\documentclass{amsart}
\usepackage{amssymb, amsmath}

\theoremstyle{plain}
\newtheorem{theorem}{Theorem}
\newtheorem{lemma}[theorem]{Lemma}

\newtheorem{cor}[theorem]{Corollary}

\theoremstyle{definition}

\renewcommand{\Re}{\operatorname{\rm Re}\nolimits}
\renewcommand{\Im}{\operatorname{\rm Im}\nolimits}

\def \comp {\operatorname{comp}}
\def \loc {\operatorname{loc}}
\def \supp {\operatorname{supp}}
\def \diam {\operatorname{diam}}

\def \Real {{\mathbb R}}

\def \Complex {\mathbb{C}}

\def \Integers {{\mathbb Z}}
\title 
[Schr\"odinger operators with no resonances]
{Schr\"odinger operators with complex-valued potentials and no resonances}
   \author { T. Christiansen}
\thanks{Partially supported by NSF grant DMS 0088922.}
\begin{document}
\begin{abstract} 
In dimension $d\geq 3$, we give examples of nontrivial, compactly supported,
complex-valued
potentials such that the associated Schr\"odinger operators have no resonances.
If $d=2$, we show that there are potentials with no resonances away from the
origin.  These Schr\"odinger operators are isophasal and have the same 
scattering phase as the Laplacian on $\Real^d$.
In odd dimensions $d\geq 3$ we study the fundamental solution
of the wave equation perturbed by such a potential.  If the space variables
are held fixed, it is super-exponentially decaying in time.
\end{abstract}

\maketitle

\section{Introduction}

In this paper we consider compactly supported, complex-valued potentials
$V \in L^{\infty}_{\comp}(\Real^d)$, $d\geq 2$.  Our first result is that
for $d\geq 3$ there are many such non-trivial $V$ so that the
meromorphic continuation of the resolvent
$(\Delta +V-\lambda^2)^{-1}$ has no poles; for $d=2$ we give examples
with no poles except, perhaps, at the origin.
These results are surprising, as there are no 
such nontrivial potentials with this property in one 
dimension
\cite{froese, regge, simon, zw1}.  Moreover, it is known that for
nontrivial
{\em real-valued}, smooth, compactly supported potentials in all dimensions
greater than two
the resolvent must have infinitely many poles
\cite{lrb,sBeven,sb-zw}.  We also show that the Schr\"odinger operators
with the potentials we
construct, like the Laplacian, have scattering phase $0$.  In addition,
in odd dimensions at least three, we show that for potentials without 
associated resonances the fundamental
solution of the perturbed wave equation is, for 
fixed values of the space variables, super-exponentially decaying in time.

Let $\Delta$ be the non-negative Laplacian on $\Real^d$, and let
$R_0(\lambda)=(\Delta-\lambda^2)^{-1}$ be the resolvent, bounded
on $L^2(\Real^d)$ for $0<\arg \lambda<\pi$.  Then, as an operator from
$L^2_{\comp}(\Real^d) $ to $H^2_{\loc}(\Real^d)$, $R_0$ has an analytic
continuation to $\Complex$ if $d$ is odd.  If $d$ is even, the continuation
is to $\Lambda$, the logarithmic cover of the complex plane (There is a 
singularity at the origin if $d=2$.).  If $V\in L^{\infty}_{\comp}(\Real^d)$,
$R_V(\lambda)=(\Delta +V-\lambda^2)^{-1}$ is bounded on $L^2(\Real^d)$
for all but a finite number of $\lambda$ with
$0<\arg \lambda<\pi$.  Like $R_0$, as an operator from $L^2_{\comp}(\Real^d)$
to $H^2_{\loc}(\Real^d)$, $R_V$ has a meromorphic continuation
to $\Complex$ (for $d$ odd) or $\Lambda$ ($d$ even).  The poles of the
continuation are called resonances, or scattering poles.  They behave 
like eigenvalues in a number of ways and may correspond to decaying states.
See \cite{vodevsurvey,zwsurvey, zwnotices,zwqr} for an introduction to 
resonances and a survey of some results on their distribution.

For $d$ odd and $V\in L^{\infty}_{\comp}(\Real^d;\Complex)$, let 
$$N_V(r)=\{ \lambda_j:\; \lambda_j \; \text{is a pole of}\; R_V
\; \text{listed with multiplicity},\;
|\lambda_j|<r\}.$$
 If $d=1$, 
$$\lim _{r\rightarrow \infty}\frac{N_V(r)}{r}=\frac{2}{\pi }\diam (\supp(V))
$$
\cite{froese, regge, simon, zw1}.  
This holds both for real and complex potentials.  If
$d\geq 3$, $N_V(r)\leq C_V(1+r^d)$, and there are radial potentials
that have resonance-counting function with this order of growth 
\cite{zwrp,zwodd}.  Lower bounds
are more delicate.  The best known lower bound to hold for a general class
of potentials is, for nontrivial $V\in C^{\infty}_{c}
(\Real^d;\Real)$,
$$\lim\sup _{r\rightarrow \infty}\frac{N_V(r)}{r}>0$$
\cite{sBlb}.  We show in this paper that it is necessary to 
assume that $V$ is real-valued,
giving evidence of the subtlety of the behaviour of resonances in dimension
bigger than one.

Upper bounds on a resonance-counting function in even dimensions can be 
found in
\cite{vodev2,vodeveven}, 
and some lower bounds for smooth real-valued potentials in
\cite{sBeven}.

On $\Real^d$ we use the ``cylindrical'' coordinates $(\rho,\theta,x')
\in [0,\infty)\times[0,2\pi)\times\Real^{d-2}$, with $x_1=\rho \cos \theta$,
$x_2=\rho\sin\theta$.  
When $d=2$, we understand that the $x'$ coordinates are omitted.  We shall
use these coordinates for the statement of the following theorem and in
Section \ref{s:pwar}.
\begin{theorem}\label{t:noresonances}
 Let $V_1\in L^{\infty}_{\comp}(\Real_+)$, $V_2\in
 L^{\infty}_{\comp}(\Real^{d-2})$, and $m\in \Integers \setminus\{0\}$.
If $d\geq 3$, let $V(\rho,\theta,x')=e^{im\theta}V_1(\rho)V_2(x')$, and
if $d=2$, let $V(\rho,\theta)=e^{im\theta}V_1(\rho)$.  Then, if $d\geq 3$,
there are no poles of $R_V(\lambda)$,
and if $d=2$, there are no poles 
of $R_V(\lambda)$ away from the origin.
\end{theorem}
These potentials are related to ones used for infinite 
cylinders in \cite{cep}.  We note that $V_1$ and $V_2$ can be chosen so 
that $V$ is smooth.

If $V_1$ and $V_2$ are real-valued, then $\Delta +V$ is a 
${\mathcal P}{\mathcal T}$-symmetric
operator \cite{c-g-s}.  Here ${\mathcal T}$ is complex-conjugation, 
$({\mathcal T}\psi)(x)=\overline{\psi}(x)$, and 
$({\mathcal P}\psi)(x_1,x_2,x')=\psi(x_1,-x_2,x')$.
See \cite{b-b,c-g-s,zn} for further references to studies of 
${\mathcal P}{\mathcal T}$-symmetric
operators, including $\mathcal{P}{\mathcal T}$-symmetric quantum mechanics and 
other applications, and \cite{l-c-v} for a study of 
scattering-theoretic questions for the one-dimensional
Schr\"odinger operator with a particular family of 
complex-valued potentials.

For a potential $W\in L^{\infty}_{\comp}(\Real^d;\Complex)$, let
$S_W(\lambda) $ be the associated scattering matrix, $s_W(\lambda)=\det S_W(\lambda)$, and $\sigma_W(\lambda)=\log s_W(\lambda)$ be the scattering phase, 
with $\sigma_W(0)\in [0,2\pi)$ to determine it uniquely.
Here $\lambda \in \Complex $ if $d$ is odd and $\lambda \in \Lambda$ if 
$d$ is even.  Like resonances, the scattering phase (for $\lambda \in \Real$)
may be thought of as an analog of discrete spectral data for our setting.
\begin{theorem}\label{t:isophasal} Let $V$ be as in Theorem 
\ref{t:noresonances}, $\lambda \in \Complex$ if $d\geq 3$ is odd,
$\lambda \in \Lambda$ if $d$ is even.  Then $s_V(\lambda)\equiv 1$.
\end{theorem}
That is, these Schr\"odinger operators 
are isophasal and have the same phase as the
Schr\"odinger operator with the 
trivial potential.  For examples of isophasal manifolds and references
to further results in that direction, see
e.g. \cite{g-p}.

In Section \ref{s:pwar} we give a direct proof of Theorem \ref{t:isophasal}
 without using 
the results of Theorem \ref{t:noresonances}. 
Here we make some comments
about the relationship between Theorems \ref{t:noresonances} and 
\ref{t:isophasal} in odd dimensions, which is the simpler case.
With at most a finite number of exceptions, the poles of $R_V$ correspond,
with multiplicity, to the poles of $s_V$.  If $R_V$ is regular at $\lambda_0$,
then so is $s_V$.  See \cite{zwpf} for a careful discussion of these questions.
Moreover $\lambda$ is a pole of $s_V$ if and only if $-\lambda$ is a zero
of $s_V$, and
the multiplicities are the same. 
Thus,
  using the Weierstrass factorization theorem,
Theorem \ref{t:noresonances},
and results of \cite{zwpf}, $s_V(\lambda)= e^{g_V(\lambda)}$,
where $g_V(\lambda)$ is a polynomial of degree at most $d$.  Further 
considerations put additional restrictions on $g_V$.  On the other hand,
if $s_V(\lambda)\equiv 1$, $R_V$ can have have at most a finite number of
poles.  A priori, one cannot rule out, for example,
 the possibility that $R_V$ has 
a pole at $\lambda_0$ and another one, of the same multiplicity, at 
$-\lambda_0$.  In this case, $s_V$ would be holomorphic at both 
$\lambda_0$ and $-\lambda_0$ \cite{zwpf}.  Thus neither 
Theorem \ref{t:noresonances} nor Theorem \ref{t:isophasal} immediately
implies the other, even in odd dimensions.

In many settings, the imaginary parts of resonances are related to the
rate of decay of solutions of a wave equation on compact sets.  In the
absence of resonances we consider the decay of the
fundamental solution $G_V(t)$ of
the perturbed wave equation
\begin{align}\label{eq:gv}
(D_t^2-(\Delta +V))G_V(t)& = 0\\
\nonumber G_V(0)& = 0\\
\nonumber (G_V)_t(0)& =I.
\end{align}  
Here $V\in L^{\infty}_{\comp}(\Real^d;\Complex)$ and 
$D_t =\frac{1}{i}\frac{\partial}{\partial t}.$  These equations uniquely
determine $G_V(t)$.
\begin{theorem} \label{t:sexpdecay}
Let $d\geq 3$ be odd,
$V\in L^{\infty}_{\comp}(\Real^d;\Complex)$, and let $G_V(t)$ be the
operator 
determined by (\ref{eq:gv}), with
$G_V(t,x,y)$ its Schwartz kernel.  Then if $R_V(\lambda)$ has no poles
for $\lambda \in \Complex$ and if $\chi \in C_c^{\infty}(\Real^d)$, then
$\chi(x)G_V(t,x,y)\chi(y)$ is super-exponentially decreasing in $t$.
\end{theorem}

As an immediate consequence of Theorems \ref{t:noresonances} and 
\ref{t:sexpdecay}, we obtain
\begin{cor}
In dimension $d\geq 3$ odd, there are nontrivial potentials 
$V\in L^{\infty}_{\comp}(\Real^d;\Complex)$ such
that the fundamental solution $G_V(t)$ determined by (\ref{eq:gv}) decays
faster than any exponential when 
the space variables are restricted to compact sets.
\end{cor}

It is a pleasure to thank N. Kalton and K. Shin for helpful conversations
 and M. Zworski for his constructive comments on an earlier version of this 
note.

\section{Proof of Theorems \ref{t:noresonances} and 
\ref{t:isophasal}}\label{s:pwar}

Recall that in this section we use the coordinates $(\rho, \theta, x')$
on $\Real^d$.
For $j\in \Integers$, let $P_j$ denote projection onto $e^{ij\theta}$.  That is,
$$(P_j f)(\rho,\theta,x')
=\frac{1}{2\pi}e^{ij\theta}\int_0^{2\pi}
f(\rho, \theta',x')e^{-ij\theta'}d\theta' .$$
\begin{lemma}
\label{l:R0Pjbd} Let $\lambda \in \Complex$ if $d$ is odd,
$d\geq 3$ and $\lambda \in 
\Lambda$ if $d$ is even.  If $d=2$, assume $\lambda \not = 0$.
For $\chi\in C_c^{\infty}(\Real^d)$ and for $j\in \Integers$, $|j|$ 
sufficiently large (depending on $\lambda$), 
$$\| \chi R_0(\lambda)P_j\chi \| \leq \frac{C}{|j|^2-C}$$
for some constant $C$ depending on $\lambda$ and $\chi$.
\end{lemma}
\begin{proof}
Note that $\Delta P_j=P_j\Delta$, so that $R_0P_j=P_jR_0$.
We may assume that $\chi$ is independent of 
$\theta$,  since the general case will follow from this special one.  Thus
we may use $\chi P_j=P_j\chi$.

We have
$$\chi(\Delta-\lambda^2)R_0(\lambda)\chi=\chi^2,$$
so that
$$(\Delta-\lambda^2)\chi R_0(\lambda)P_j\chi = \chi^2P_j -[\chi,\Delta]
R_0(\lambda)\chi P_j.$$
Then
\begin{align*}
\|(\Delta-\lambda^2)\chi R_0(\lambda)P_j\chi\|&  \leq 
\|\chi^2P_j\|+\|[\chi,\Delta]R_0(\lambda)\chi P_j\|\\
& \leq C
\end{align*}
where the constant depends on $\lambda $ and $\chi$.
But, for some $c_{\chi}>0$,
$$\|(\Delta-\lambda^2)\chi P_j v\| \geq (c_{\chi}j^2-\Re \lambda^2)\|
\chi P_j v\|.$$
Thus, when $|j|$ is so large that 
$c_{\chi}j^2-\Re \lambda^2>0$, 
$$\| \chi R_0(\lambda)\chi P_j\| \leq \frac{C}{c_{\chi}j^2-\Re \lambda^2}.$$
\end{proof}

We shall use the following elementary lemma in our proof of the theorem.
\begin{lemma}\label{l:l2}
Let $\{a_j\}_{j=-\infty}^{\infty}\in \ell^2$ and $m\in \Integers$, 
$m\not = 0$. 
 Suppose for each $j$ there
is a constant $C_j$ such that $|a_{j+m}|\leq C_j |a_j|$.  If, in
addition, there is a $J$ 
 such that $|a_{j+m}|\leq |a_j|$ when $|j|\geq J$, then $a_j=0$ for all $j$.
\end{lemma}
\begin{proof} 
Suppose $a_{j_0}\not = 0$ for some $j_0$.  Then since 
$|a_{j+m}|\leq C_j |a_j|$,
$a_{j_0-km}\not = 0
$ for $k=1,\;2,\;3,....$  

Since $\{a_j\}\in \ell^2$, $|a_{j_0-km}|\rightarrow 
0$ as $k\rightarrow \infty$.  But when $|j_0-(k+1)m|>J$,
$$|a_{j_0-km}|\leq |a_{j_0-(k+1)m}|\leq |a_{j_0-(k+2)m}|\leq ....$$
Thus $a_{j_0-km}=0$, a contradiction.
\end{proof}

Now we are able to give the proof of the first theorem.
\begin{proof}[Proof of Theorem \ref{t:noresonances}]
Suppose, on the contrary, $\lambda\in \Complex $ 
if $d$ is odd or $\lambda \in \Lambda$ if $d$ is
even is a pole of the resolvent, and $\lambda$ is not the origin if
$d=2$.  Then, since
$$(\Delta+V-\lambda^2)R_0(\lambda)=I+VR_0(\lambda)$$
and $R_0$ is holomorphic (if $d\geq 3$), or holomorphic away from the 
origin (if $d=2$), 
 there is a nontrivial $u\in L^2(\Real^d)$ so that 
$$(I+VR_0(\lambda))u=0.$$
The function $u$ is necessarily supported on $\supp V$.  We can write
$$u(\rho, \theta, x')=\sum_{-\infty}^{\infty}u_j(\rho, x')e^{ij\theta}$$
(where we omit the $x'$ variables if $d=2$). 
Since
$$ u=-VR_0(\lambda)u,$$
if $\chi\in C_c^{\infty}(\Real^d)$, with $\chi\equiv 1$ on 
$\supp V$, 
\begin{equation*}
u_{j+m}=-V_1V_2\chi R_0(\lambda)\chi P_j u 
\end{equation*}
and
$$\|u_{j+m}\|_{L^2(\Real^d)}\leq C_j\|u_j\|_{L^2(\Real^d)}.$$
By Lemma \ref{l:R0Pjbd}, when $|j|$ is sufficiently large,
$$\|u_{j+m}\|_{L^2(\Real^d)}\leq \frac{C}{j^2-C}\|u_j\|_{L^2(\Real^d)}.$$
Using Lemma \ref{l:l2} applied to
$\{ \|u_j\|_{L^2(\Real^d)}\}$, $\|u_j\|_{L^2(\Real^d)}=0$ 
for all $j$, and thus $u\equiv 0$.
\end{proof}

The proof of Theorem \ref{t:isophasal} does not use any of the other 
results of this section.
\begin{proof}[Proof of Theorem \ref{t:isophasal}]
Fix $\lambda \in  \Complex$ if $d$ is
odd and $\lambda \in \Lambda$ if $d$ is even.  If 
$d=2$ assume that $\lambda$ is not the origin.  Recall that for 
$W_{\comp}^{\infty}(\Real^d;\Complex)$, $S_W$ denotes the scattering 
matrix associated with
$\Delta +W$ and $s_W(\lambda)=\det S_W(\lambda)$.

Let $V=V(\rho,\theta,x')$ be as defined in Theorem \ref{t:noresonances}
and, for $z\in \Complex$, let $W_z(\rho,\theta,x')=z^mV(\rho,\theta,x').$ 
If $m>0$, $W$ is holomorphic as a function of $z\in \Complex$, and 
if $m<0$, it is holomorphic
on $\Complex \setminus \{ 0\}$, and meromorphic on $\Complex$.  Then, for 
fixed $\lambda$, $S_{W_z}(\lambda)$
and 
$$h_{\lambda}(z)=\det S_{W_z}(\lambda)$$ 
depend meromorphically on $z\in \Complex$.

For $\phi \in \Real$, $(e^{i\phi})^mV(\rho,\theta,x')
= V(\rho,\theta+\phi,x')$ where we make the identification
$\theta+2k\pi=\theta$ for $k\in \Integers$.  That is, 
$(e^{i\phi})^mV$ corresponds to $V$ under a rotation of angle
$\phi$ in the $x_1x_2$ plane.  Although in general the scattering matrix
$S_V(\lambda)$
is not invariant under rotations of $\Real^d$, its eigenvalues are, and 
so 
$s_V(\lambda)$ is unchanged.
Thus
$$h_{\lambda}(e^{i\phi})=s_{e^{i\phi}V}(\lambda)=s_V(\lambda)=h_{\lambda}(1).$$
Then $h_{\lambda}(z)$ is a meromorphic function of $z\in \Complex$ which is 
constant on the unit circle.  Thus it is a constant function
$h_{\lambda}(z)\equiv h_{\lambda}(1).$

If $m>0$, 
$$h_{\lambda}(1)= h_{\lambda}(z)= h_{\lambda}(0) = \det S_0(\lambda)=1.$$
If $m<0$,
$$h_{\lambda}(1)= h_{\lambda}(z)= 
\lim_{r\rightarrow \infty}h_{\lambda}(r) = \lim_{r\rightarrow \infty}
\det S_{r^m V}(\lambda)=1.$$

This proves the theorem except for the question of what happens at the 
origin if $d=2$.
However,
 the regularity properties of $s_V$ imply that it is $1$
at the origin as well.
\end{proof}

\section{Proof of Theorem \ref{t:sexpdecay}}

In this section, we show that for odd $d$ 
and for $V\in L^{\infty}_{\comp}(\Real^d;\Complex)$
without resonances the fundamental solution of the perturbed
wave equation $D^2_t-(\Delta +V)$ decays super-exponentially.

\begin{lemma}\label{l:specrep}
For $V\in L^{\infty}_{\comp}(\Real^d;\Complex)$ such that $\Delta +V$ 
has no eigenvalues,
\begin{equation}\label{eq:specrep}
G_V(t)=\frac{1}{2\pi i}
\int _{-\infty}^{\infty} \sin (t\lambda) \left( R_V(\lambda)-R_V(-\lambda)
\right)d\lambda \end{equation}
where $G_V(t)$ is determined by (\ref{eq:gv}).
\end{lemma} 
If $V$ is real-valued, this lemma follows immediately from
the spectral theorem and
Stone's formula.  We are unaware of a reference 
that would directly imply this result for complex-valued potentials
and thus include a proof below.
\begin{proof}  
Let $f\in C_c^{\infty}(\Real^d)$ and
\begin{equation}\label{eq:uf}
u_f(t)=\frac{1}{2\pi i}
\int _{-\infty}^{\infty} \sin (t\lambda) \left( R_V(\lambda)-R_V(-\lambda)
\right)fd\lambda.
\end{equation}  We shall show that 
\begin{align}\label{eq:ufe}
(D_t^2-(\Delta +V))u_f(t)& = 0\\
\nonumber u_f(0)& = 0\\
\nonumber (u_f)_t(0)& =f.
\end{align}  This will show that the operator determined by 
the right-hand side of (\ref{eq:specrep}) agrees with $G_V(t)$ on a dense
subspace of $L^2(\Real^d)$, and thus, since $G_V(t)$ is
continuous, the two coincide.  Here we 
use the uniqueness of solutions of the initial value problem for the
perturbed wave equation.

We use the notation $\langle x\rangle = (1+|x|^2)^{1/2}$.
When $\Im \lambda \geq 0$ and $s>1/2$
$$\| \langle x \rangle^{-s} R_0(\lambda)\langle x \rangle^{-s}\| 
\leq \frac{C}{|\lambda|}$$
\cite[Corollary 3.6]{yafaev}
and, since $\Delta +V$ has no eigenvalues,
$$
\| \langle x \rangle^{-s} R_V(\lambda)\langle x \rangle ^{-s}\| \leq \frac{C}{|\lambda|}$$
for the same $\lambda$ and $s$.
  Since 
$$\left(R_V(\lambda)-R_V(-\lambda)
\right)f= \frac{1}{\lambda^2}\left(R_V(\lambda)-R_V(-\lambda)
\right)(\Delta+V)f$$
the integral in (\ref{eq:uf}) converges absolutely as 
an element of $\langle x \rangle ^s L^2(\Real^d)$, for any $s>1/2$.  It 
is easy to see then that the first two equalities in (\ref{eq:ufe}) hold.

To prove the third equality in (\ref{eq:ufe}) we use the $H_{\infty}$ 
functional calculus \cite{mcintosh}.  Thus
\begin{equation} 
f  =\frac{1}{2\pi i} \int_{\gamma} \frac{1}{\mu +10}
(\Delta +V-\mu)^{-1}((\Delta +V+10)f)d\mu
\end{equation}
where 
$\gamma$ is the contour defined by 
the function
\begin{equation}
g(t)=\left\{\begin{array}{ll}
-te^{-i\theta}-1,& t\leq 0 \\
te^{i\theta}-1, & t\geq 0
\end{array}\right.
\end{equation}
and $\theta$ is chosen sufficiently small that $\gamma$ does not
enclose $\pm i \sqrt{10}$.  Now we shall do some contour deformation
which is valid because $(\Delta +V-\mu)^{-1}$ has no poles
and because the integrand (as an element of $\langle x \rangle^sL^2(\Real^d)$,
$s>1/2$)
is bounded in norm by $C\mu^{-3/2}$ when $|\mu|$ is large.   Doing a contour
integration and a change of variables, we see that 
$$f=\frac{1}{\pi i} \int _{\Im \lambda =\epsilon>0}
\frac{1}{\lambda^2+10}R_V(\lambda)((\Delta+V+10)f)\lambda d\lambda.$$
Taking the limit as $\epsilon \downarrow 0$, a
simple further change of variables, and noting that
$(R_V(\lambda)-R_V(-\lambda))((\Delta+V+10)f)=(\lambda^2+10)
(R_V(\lambda)-R_V(-\lambda))f$
finishes the 
proof.

\end{proof}

Using this lemma, the proof of Theorem \ref{t:sexpdecay} is straightforward.
\begin{proof}[Proof of Theorem \ref{t:sexpdecay}]
Recall the resolvent equation:
\begin{equation}
R_V(\lambda)=R_0(\lambda)-R_V(\lambda)VR_0(\lambda).
\end{equation}
Thus, using Lemma \ref{l:specrep}, 
$$G_V(t)=G_0(t)-\frac{1}{2\pi i}
\int _{-\infty}^{\infty} \sin (t\lambda) \left( R_V(\lambda)VR_0(\lambda)
-R_V(-\lambda)VR_0(-\lambda)
\right)d\lambda.$$
Since $G_0(t,x,y)$ is supported on $|x-y|=|t|$, it
is super-exponentially decaying on compact sets in $x$ and $y$.

Let $\chi' \in C_c^{\infty}(\Real^d)$.  Then 
$$\|\chi' R_0(\lambda)\chi' \|
\leq \frac{C}{1+|\lambda|}(1+e^{-C\Im \lambda})$$
\cite[(16)]{zwodd}.
Using this, for $M\in \Real$,
$\Im \lambda>-M$, $|\lambda|$ sufficiently large,
$$R_V(\lambda)\chi=R_0(\lambda)\sum_0^{\infty}(-1)^j(VR_0(\lambda))^j\chi,$$
recalling that $\chi \in C_c^{\infty}(\Real^d)$.
Thus, using in addition the fact that $R_V$ has no poles,
$$\|\chi R_V(\lambda)\chi \| \leq \frac{C}{1+|\lambda|}(1+e^{-C\Im \lambda})$$
when $\Im \lambda>-M$. 
These estimates, along with the fact that $R_V$ has no poles, allows us
to see by contour deformation 
\begin{align*} &
 \int _{-\infty}^{\infty} e^{it\lambda} \chi \left( R_V(\lambda)VR_0(\lambda)
-R_V(-\lambda)VR_0(-\lambda)
\right)\chi d\lambda \\ &
=  \int _{-\infty}^{\infty} e^{it(\lambda+ i\beta)} 
\chi \left( R_V(\lambda+i\beta)VR_0(\lambda+i\beta)
-R_V(-\lambda-i\beta)VR_0(-\lambda-i\beta)
\right) \chi d\lambda \\&
= O(e^{-t\beta})
\end{align*} for $\beta \in \Real.$
A similar argument shows that
$$\int _{-\infty}^{\infty} e^{-it\lambda} \chi \left( R_V(\lambda)VR_0(\lambda)
-R_V(-\lambda)VR_0(-\lambda)
\right)\chi d\lambda  =O(e^{-t\beta}).$$
Since $\beta$ is arbitrary,  $\chi (x) G_V(t,x,x')\chi(x')$ decays
faster than any exponential.
\end{proof}

\small
\noindent
{\sc 
Department of Mathematics\\
University of Missouri\\
Columbia, Missouri 65211\\
{\tt tjc@math.missouri.edu}}
\end{document}